\documentclass[10pt, devips]{article}
\usepackage{float}
\usepackage{graphicx}
\usepackage{amsmath}
\usepackage{amssymb}
\usepackage{latexsym}
\usepackage{color}
\usepackage{graphicx}
\usepackage{caption}
\usepackage{xcolor} 
\usepackage[colorlinks=red,linkcolor=red]{hyperref}

\usepackage{subfig}
\usepackage{cite}
\usepackage{romanbar}
\usepackage[squaren,Gray]{SIunits}
\begin{document}
	\thispagestyle{empty}
	\begin{center}
		
		\vspace{1.8cm}
		
		{\large {\bf Tunable phonon-photon coupling induces double MMIT and enhances slow light in an atom-opto-magnomechanics}}\\

		\vspace{1.5cm}
		
		{\bf M'bark Amghar}$^{}${\footnote { email: {\sf
					amghar.mbark98@gmail.com}}}, {\bf Noura Chabar}$^{}${\footnote { email: {\sf
					chabarnoura@gmail.com}}} and {\bf Mohamed Amazioug}$^{}${\footnote { email: {\sf
					amazioug@gmail.com}}}
		
		\vspace{0.5cm}
		
		{\it LPTHE, Department of Physics, Faculty of Sciences, Ibnou Zohr University,\\  Agadir, Morocco.}

		\vspace{1.5cm} {\bf Abstract}
	\end{center}
	In this paper we theoretically investigate the magnomechanically induced transparency phenomenon and the slow/fast light effect in the situation where an atomic ensemble is placed inside the hybrid cavity of an opto-magnomechanical system. The system is driven by dual optical and phononic drives. We show double magnomechanically induced transparency (MMIT) in the probe output spectrum by exploiting the phonon-photon coupling strength. In addition, the fast and slow light effects in the system are explored. Besides, we show that the slow light profiles is enhanced by adjusting phonon-photon coupling strength. This result may have potential applications in quantum information processing and communication.

	\vspace{0.25cm}
	\textbf{Keywords}: Cavity magnomechanics, Magnomechanically induced transparency, Absorption, Dispersion, Transmission, Slow light, Fast light.

	\section{INTRODUCTION}
	
	The interaction between light and matter has emerged as a dynamic and significant area of research, with broad potential applications across various domains. These applications include optomechanically induced transparency (OMIT) \cite{a0,a1,a2,a3,mas}, coherent control of light \cite{a4,a5}, electromagnetically induced transparency (EIT)\cite{1,2,3,4,5,6}, and the preparation of entanglement between cavity fields and macroscopic oscillators \cite{ma1,ma2,9,7,ma3}. Recently, this research has played a pivotal role in advancing the field of solid-state quantum memory and quantum information processing. EIT, or Electromagnetically Induced Transparency, is a common quantum interference effect observed in three-level atoms \cite{10}. It describes the intriguing phenomenon where the absorption rate of atoms can be diminished to zero when exposed to an auxiliary laser field. This reduction is primarily a result of interference effects or dark state resonance in an excited state. Furthermore, analogous to EIT, when destructive interference occurs between the anti-Stokes scattering field and the weak probe field, it is referred to as OMIT \cite{11}. This phenomenon has been the subject of extensive research, both theoretically \cite{a0,a1} and experimentally \cite{a2,a3}. The multiple-OMIT phenomena has been studied theoretically in a variety of innovative cavity optomechanical systems that have emerged in recent years, including hybrid piezooptomechanical cavity systems \cite{12}, atomic-media assisted optomechanical systems \cite{13}, multiple-resonator optomechanical systems \cite{14}, and others.\\

	Similar to cavity optomechanics (COM), cavity magnomechanics (CMM) \cite{15,16,17,18} has drawn more attention lately due to its potential for larger-scale quantum state preparation \cite{19,20,21} and its numerous intriguing uses in quantum information science and technologies \cite{22,23,24,25,26,27,28}. It investigates interactions between microwave cavity photons, magnons (quanta of the spin wave), and magnetostriction-induced vibration phonons in magnetically ordered materials, such as yttrium-iron-garnet (YIG) \cite{15,16,17,18}. The combination of cavity optomechanics and cavity magnomechanics, realized by coupling the magnomechanical displacement to an optical cavity via radiation pressure, creates the new system of opto-magnomechanics (OMM) \cite{29,30,31}. This hybrid system provides the capability to optically detect the population of magnons \cite{29}, manipulate diverse magnonic quantum states within solid materials \cite{30}, and generate entanglement between optomagnonic \cite{30} and microwave-optics \cite{31}. Consequently, this system holds great promise for applications in quantum information processing and the development of quantum networks. Recently, the magnomechanically induced transparency, and slow-fast light has been studied in magnomechanical system by modifying the strength of the atom-cavity coupling, without taking into consideration the cavity-phonon coupling \cite{311}. \\

	In this work, we propose and analyze the opto-magnomechanically induced transparency phenomenon and fast-slow light effect in an opto-magnomechanical system in which a high-quality yttrium iron garnet (YIG) sphere and an atomic ensemble are contained within a microwave cavity (Fig. \eqref{701}). Magnon-phonon interactions give rise to magnomechanically induced transparency (MMIT) and photon-phonon interaction (OMIT) in the probe output spectrum. Furthermore, we explore the impact of optomechanical coupling on the absorption spectrum. In addition, we discuss slow-light propagation phenomena. We demonstrate that, in relation to the magnomechanical coupling of the YIG and the atom-cavity coupling strength, the group delay is dependent on the tunability of the optomechanical coupling strength.  \\

	The structure of the paper is organized as follows. In Sect. \ref{one}, we present the formula of the Hamiltonian and the appropriate quantum Langevin equations of the system under considerationand obtain the output field. In sect. \ref{two}, we discuss the magnomechanically induced transparency and analyze the influence of the optomechanical coupling strength on the input spectrum. In Sec. \ref{tree}, we describe the probe field transmission and discuss the group delays for slow and fast light propagation. Concluding remarks close this paper.
	
	\section{MODEL }\label{one}
	
	\subsection{Hamiltonian}
	
	\begin{figure}[t]
		\centering
		\hskip-1.0cm\includegraphics[width=0.7\linewidth]{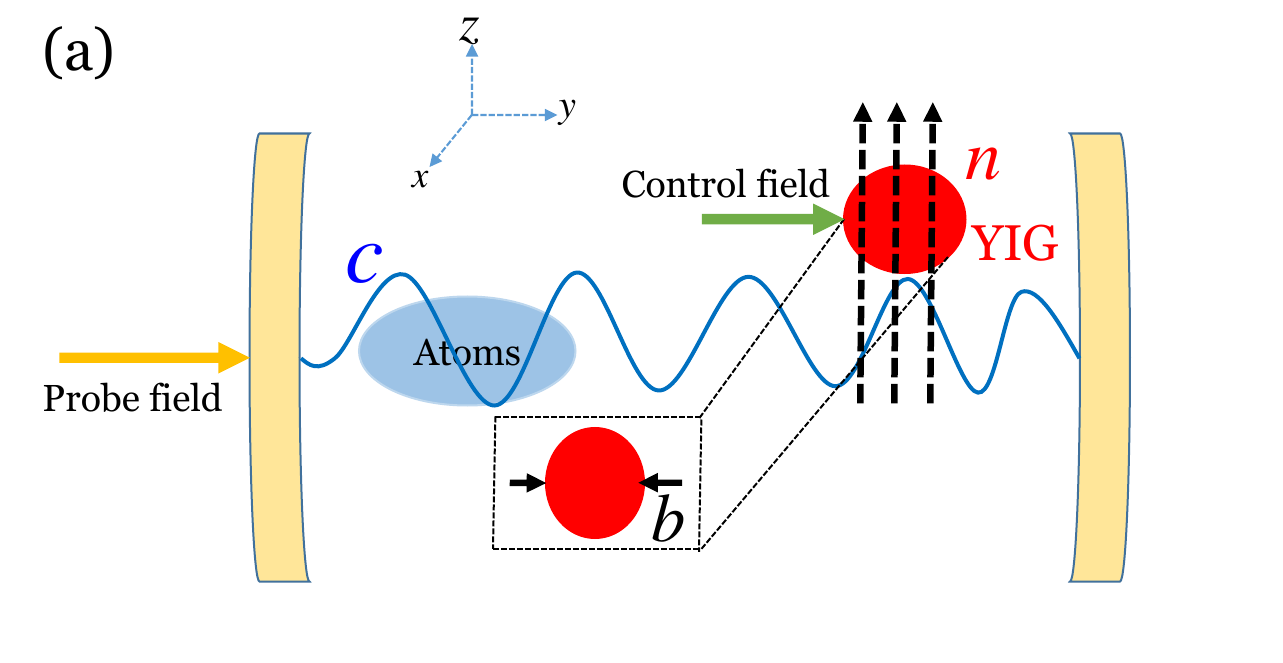}
		\hskip-1.0cm\includegraphics[width=0.6\linewidth]{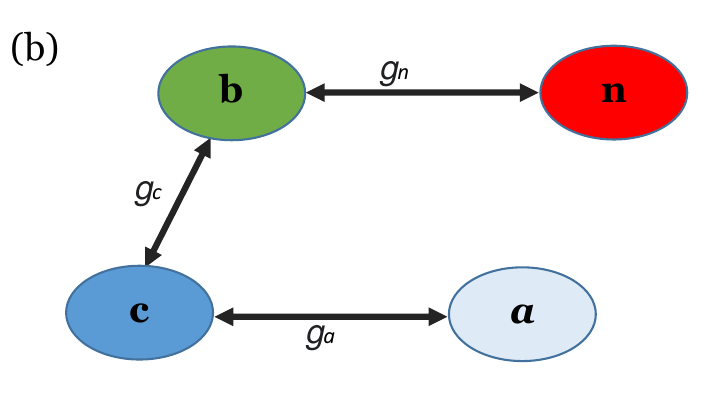}
		\caption{(a) Schematic diagram of the atom-opto-magnomechanical system. Through the mediating effect of a mechanical vibration mode $b$ induced by magnetostriction, optical cavity mode $c$, driven by a probe field at frequency $\omega_p$, couples to an ensemble of two-level atoms $a$ and magnon mode $n$ in a YIG crystal. (b) The different coupling strengths in our system are: $g_c$ denotes the optomechanical coupling strength, $g_n$ is the magnomechanical coupling strength, and $g_a$ is the atom-cavity coupling strength.  }\label{701}
	\end{figure}
	
In this subsection, we consider a hybrid cavity opto-magnomechanical system \cite{15}, composed of optical cavity photons, magnetostriction-induced vibration phonons, magnons in a YIG crystal, and an ensemble of two-level atoms, as shown in Fig. \ref{701}(a). In the system, a uniform bias magnetic field is applied to the sphere YIG in the z-direction, exciting the magnon modes. These modes are coupled with the photon mode of the cavity through magnetic dipole interactions \cite{32}. When magnon modes are excited inside spheres, variable magnetization occurs, resulting in the deformation of their lattice structure. The phonon mode results from the vibrations of the YIG sphere due to the magnetostrictive force. Hence, the coupling between magnons and phonons was established within the YIG sphere. \cite{15}. The Hamiltonian of the system is given by \cite{b1}  
	\begin{equation}\label{e1}
		\begin{aligned}
			\mathcal{H} = \mathcal{H}_{\text{free}}+\mathcal{H}_{\text{int}}+\mathcal{H}_{\text{drive}},
		\end{aligned}
	\end{equation}
	where the first part of the $\mathcal{H}$ is the free Hamiltonian terms
	\begin{equation}
		\begin{aligned}
			\mathcal{H_\text{free}} =\hbar \left( \omega_c c^{\dagger} c+\omega_n n^{\dagger} n+\frac{\omega_a}{2} {S}_z+\frac{\omega_b}{2}\left(x^2+y^2\right)\right),
		\end{aligned}
	\end{equation}
	 the first and second terms are respectively describes the energy of the optical and magnon modes with $c (c^\dagger)$ and $n (n^\dagger)$ are the annihilation (creation) operators satisfying the usual commutation relation $[\mathcal{A}, \mathcal{A}^\dagger] = 1$ where $\mathcal{A} = c,n$. The parameters $\omega_c$ and $\omega_n$ denote the resonance frequencies of the cavity and magnon modes, respectively. It's worth noting that the value of frequency $\omega_m$ can be adjusted by altering the external bias magnetic field $B$. The third term gives the energy of the atomic ensemble comprised of $N_a$ two-level atoms with natural frequency $\omega_a$. ${S}_+$, ${S}_-$ and ${S}_z$ represent the spin operators of the atoms, defined as ${S}_{+,-,z} =\sum_{i}\sigma^{(i)}_{+,-,z} $, with $\sigma_+$, $\sigma_-$ and  $\sigma_z$ are Pauli matrices, and satisfy the commutation relations $[{S}_+, {S}_-] = {S}_z$ and $[{S}_z, S_{\pm}] = \pm2{S}_{\pm}$. The fourth term describes the energy of the mechanical vibration mode with $x$ and $y$ being respectively the position and momentum operator such that $[x, y] = i$. $\omega_b$ is the resonance frequency of the mechanical mode. The second part of the $\mathcal{H}$ is the interaction Hamiltonian
	\begin{equation}
		\begin{aligned}
			\mathcal{H_\text{int}} = \hbar \left(-g_c c^{\dagger} c q+g_n n^{\dagger} n q+ g_a\left({S}_{+} c+{S}_{-} c^{\dagger}\right)\right),
		\end{aligned}
	\end{equation}
	the first term describes the interaction between the mechanical mode and the optical mode of rate $g_c$. The second term represents the interaction between the mechanical mode and the magnon mode of rate $g_n$. The last term corresponds to the interaction between atomic mode and optical mode of rate $g_a=\rho \sqrt{\omega_c/2\hbar\epsilon V}$, with $\rho$ as the atomic dipole moment, $\epsilon$ is the vacuum permittivity, and $V$ is the volume of the cavity. The last part of the $\mathcal{H}$ is the driving Hamiltonian
	\begin{equation}
		\begin{aligned}
			\mathcal{H_\text{drive}} =  i( c^{+}\epsilon_p e^{-i\omega_pt}-\text{H.c})+i(\Omega_Ln^{+}e^{-i\omega_Lt}-\text{H.c}),
		\end{aligned}
	\end{equation}
	 where $\epsilon_p=\sqrt{2\kappa_c\mathcal{P}_p/\hbar \omega_p}$ being the coupling strength between the cavity and the probe field, with $\omega_p$ ($\mathcal{P}_p$) represents the frequency (power) of a probe field and $\kappa_c$ is decay rate of the cavity. The Rabi frequency $\Omega_L=\frac{\sqrt{5}}{4}\gamma\sqrt{N}B_L$ being the coupling strength between the magnon mode and the drive field with amplitude $B_L$ and frequency $\omega_L$. $\gamma$ is the gyromagnetic ratio and $N$ is the total number of spins in the YIG. If we assume that all atoms are initially in the ground state, so that $S_z  \simeq  \langle S_z\rangle \simeq  -N_a$, and the excitation probability of a single atom is small, we can define the dynamics of the atomic polarization by bosonic annihilation operators $a = S_-/ \sqrt{|\langle S_z\rangle|}$, and satisfy the commutation relation $[a, a^\dagger] = 1$. So, we obtain the total bosonized Hamiltonian, given by
	\begin{equation}\label{1}
		\begin{aligned}
			\mathcal{H} / \hbar= & \omega_c c^{\dagger}c+\omega_n n^{\dagger} n+\omega_a a^{\dagger} a+\frac{\omega_b}{2}\left(x^2+y^2\right) 
			+g_N\left(a^{\dagger} c+a c^{\dagger}\right)-g_c c^{\dagger} c q+g_n n^{\dagger} n q\\
			&+i(\Omega_Ln^{+}e^{-i\omega_Lt}-\text{H.c}) +i( c^{+}\epsilon_p e^{-i\omega_pt}-\text{H.c}),
		\end{aligned}
	\end{equation}
	where $g_N = g_a \sqrt{N_a} $ is the effective atom-cavity coupling strength.
	
\subsection{Quantum Langevin equations and output field}
	
	The Heisenberg Langevin equations below can be used to characterize the system's quantum dynamics
	\begin{equation}\label{x}
		\begin{aligned}
			\dot{a} & =-i \Delta_a a-\gamma_a a-i g_N c+\sqrt{2 \gamma_a} a_{\mathrm{in}}, \\
			\dot{c} & =-i \Delta_c c-\kappa_c c+i g_c c q-i g_N a+\epsilon_p e^{-i\delta t}+\sqrt{2 \kappa_c} c_{\mathrm{in}}, \\
			\dot{n} & =-i \Delta_n n-\kappa_n n-i g_n n q+\Omega_L+\sqrt{2 \kappa_n} n_{\mathrm{in}}, \\
			\dot{q} & =\omega_b p, \quad \dot{p}=-\omega_b q-\gamma_b p+g_c c^{\dagger} c-g_n n^{\dagger} n+\xi,
		\end{aligned}
	\end{equation}
	where $\delta=\omega_p-\omega_L$, $\Delta_{c}=\omega_{c}-\omega_L$, $\Delta_{n}=\omega_{n}-\omega_L$ and $\Delta_{a}=\omega_{a}-\omega_L$ are the detuning parameters. $\gamma_b$ and $\kappa_n$ are the dissipation rate of the mechanical and magnon modes, respectively. $\gamma_a$ is the decay rate of the atoms. The terms $a_{in}(t)$, $c_{in}(t)$ and $n_{in}(t)$ are the vacuum input noise operators, which have zero-mean values. $\xi(t)$ being the Hermitian Brownian noise operator acting on the mechanical oscillator.\\\vspace{0cm}
	\hspace{0.5cm}The quantum Langevin equations \ref{x} can be linearized by focusing on the steady-state values and only using first-order terms in the fluctuating operator: $\langle \mathcal{X}\rangle=\mathcal{X}_{0}+\mathcal{X}_{-} e^{-i \delta t} +\mathcal{X}_{+}e^{i \delta t}$, where $\mathcal{X}= a, c, n, q, p.$ First, we examine the zero-order solution, i.e. the steady-state solutions, represented as follows: 
	\begin{equation}
		\begin{gathered}
			a_0=\frac{-i g_N c_0}{\gamma_a+i \Delta_a} \\
			c_0=\frac{-i g_N a_0}{\kappa_c+i\left(\Delta_c-g_c q_0\right)}=\frac{-i g_N a_0}{\kappa_c+i \bar{\Delta}_c} \\
			n_0=\frac{\Omega_L}{\kappa_n+i\left(\Delta_n+g_n q_0\right)}=\frac{\Omega_L}{\kappa_n+i \bar{\Delta}_n} \\
			q_0=\frac{g_c\left|c_0\right|^2-g_n\left|n_0\right|^2}{\omega_b},
		\end{gathered}
	\end{equation}
	where $\bar{\Delta}_c=\Delta_c-g_c q_0$ and $\bar{\Delta}_n=\Delta_n+g_n q_0$ are the effective detunings of the cavity and the magnon, respectively. We suppose that the coupling of the external microwave drive on magnon mode $n$ is much stronger than the coupling of the external probe field $\epsilon_p$. Under this assumption, the solution of linearized quantum Langevin equations can be considered to be the first-order external probe field by ignoring all higher-order terms of $\epsilon_p$. So, the solution for the cavity mode is  given as,
	\begin{equation}
		c_-=\epsilon_p\left( h_3+\frac{g_N^2}{h_1}-\frac{G_{c}}{\sqrt{2}}\frac{\mathcal{{H}}_6}{\mathcal{H}_5}\right) 
	\end{equation}
	Where
	$$h_1=\gamma_a+i(\Delta_a-\delta),\quad h_2=\gamma_a+i(\Delta_a+\delta) ,\quad h_3=\kappa_c+i(\bar{\Delta}_c-\delta),\quad h_4=\kappa_c+i(\bar{\Delta}_c+\delta)$$
	$$h_5=\kappa_n+i(\bar{\Delta}_n-\delta),\quad h_6=\kappa_n+i(\bar{\Delta}_n+\delta) ,\quad h_7=\omega_b^2-\delta^2-i\gamma_b\delta,\quad h_8=\omega_b^2-\delta^2+i\gamma_b\delta$$
	$$\mathcal{H}_1=1+\frac{g_N^2}{h_2h_4},\quad \mathcal{H}_2=ih_8^*-\omega_b\left( \frac{G_{c}^2}{2\mathcal{H}_1^*h_4^*}+\frac{G_n^2}{2h_6^*}\right) ,\quad \mathcal{H}_3=\omega_b\frac{-G_n^2}{2h_5}-ih_7,\quad \mathcal{H}_4=\omega_b\left( \frac{G_{c}^2}{h_4^*\mathcal{H}_1}+\frac{G_n^2}{2h_6^*}\right) $$
	$$\mathcal{H}_5=\mathcal{H}_3-\frac{\omega_b G_n^2\mathcal{H}_4}{2h_5\mathcal{H}_2},\quad \mathcal{H}_6=\omega_b\frac{G_{c}}{\sqrt{2}}\left( 1+\frac{\mathcal{H}_4}{\mathcal{H}_2}\right) $$
	Here, $G_c = i\sqrt{2}g_cc_0$ and $G_n = i\sqrt{2}g_nn_0$ are the effective opto- and magnomechanical coupling strengths. We use the input-output standard relation 
	$\epsilon_{\text{out}}=\epsilon_{\text{in}}-2\kappa_c\langle c \rangle $ \cite{b2}, where $\epsilon_{\text{in}}$ ($\epsilon_{\text{out}}$) being the input field (output field) vector, and the amplitude of the output field can be written as
	\begin{equation}\label{0}
		\epsilon_{\text{out}}=\frac{2\kappa_cc_-}{\epsilon_p}=\epsilon_{\text{R}}+i\epsilon_{\text{I}},
	\end{equation}
	where, $\epsilon_{\text{R}}$ is the real part of $\epsilon_{\text{out}}$, which describes the absorption spectrum of the output probe field frequency.  $\epsilon_{\text{I}}$ is the imaginary part of $\epsilon_{\text{out}}$, which describes the dispersion spectrum of the probe field frequency.
	
	\section{MAGNOMECHANICALLY INDUCED TRANSPARENCY}\label{two}
	
	\hspace{0.5cm}In this section, we will study the magnomechanically induced transparency in the system under consideration and analyze the absorption and dispersion spectra of the output field. The parameters we use were obtained in a recent experiment on a hybrid magnomechanical system \cite{311,15,16}: $\omega_n/2\pi=10$ GHz, $\omega_b/2\pi=40$ MHz, $\kappa_n/2\pi=1$ MHz, $\gamma_b=\kappa_n$, $\kappa_c=2\kappa_n$, $\gamma_b/2\pi=10^2$ Hz, $g_N/2\pi = 8$ MHz, $\Delta_a=-\omega_b$, $\bar{\Delta}_c=0.5\omega_b$.
	\begin{figure} [h!] 
		\begin{center}
			\includegraphics[scale=0.35]{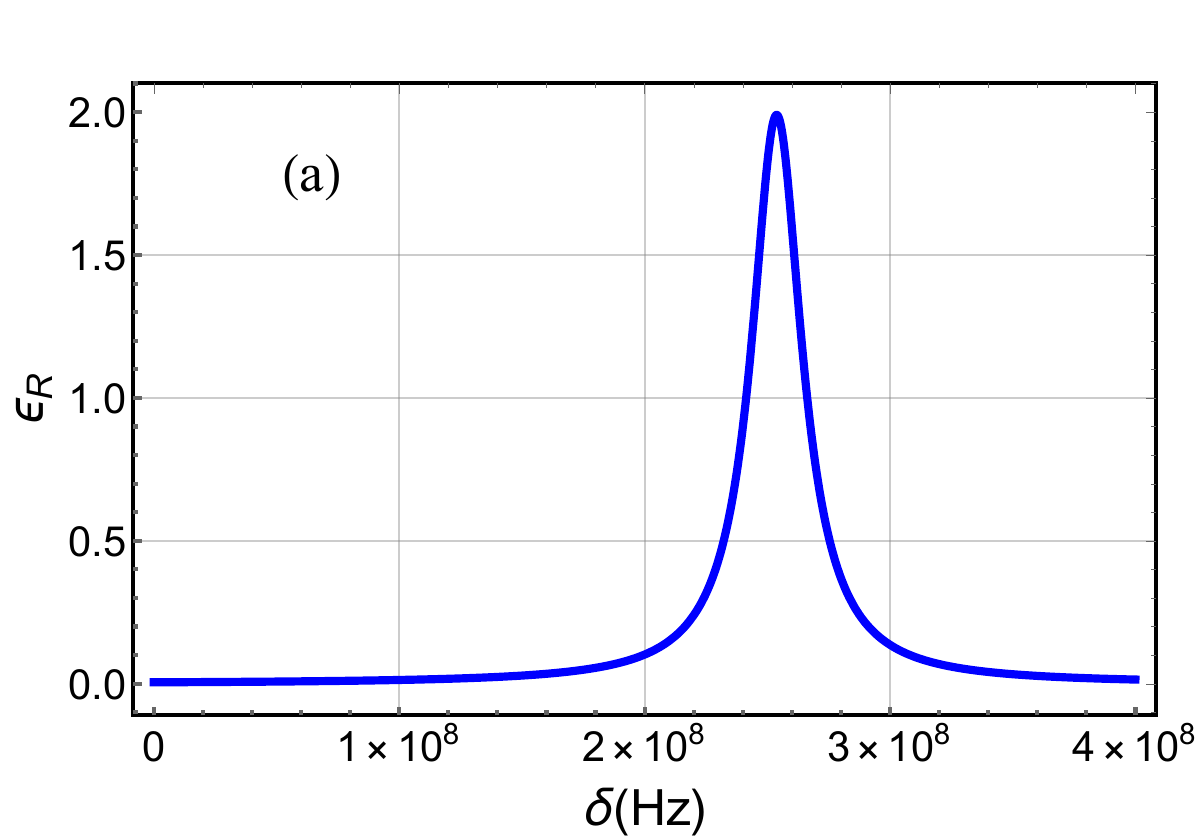}
			\includegraphics[scale=0.35]{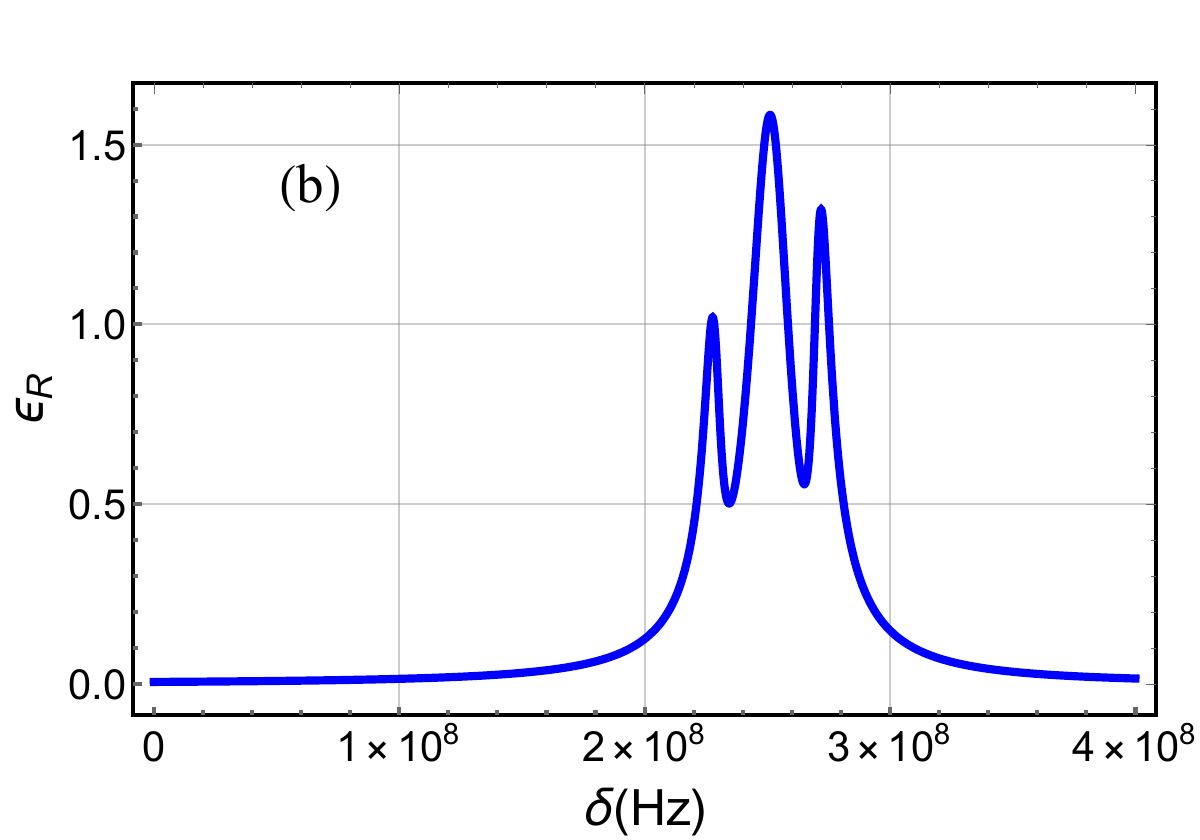}
			\includegraphics[scale=0.35]{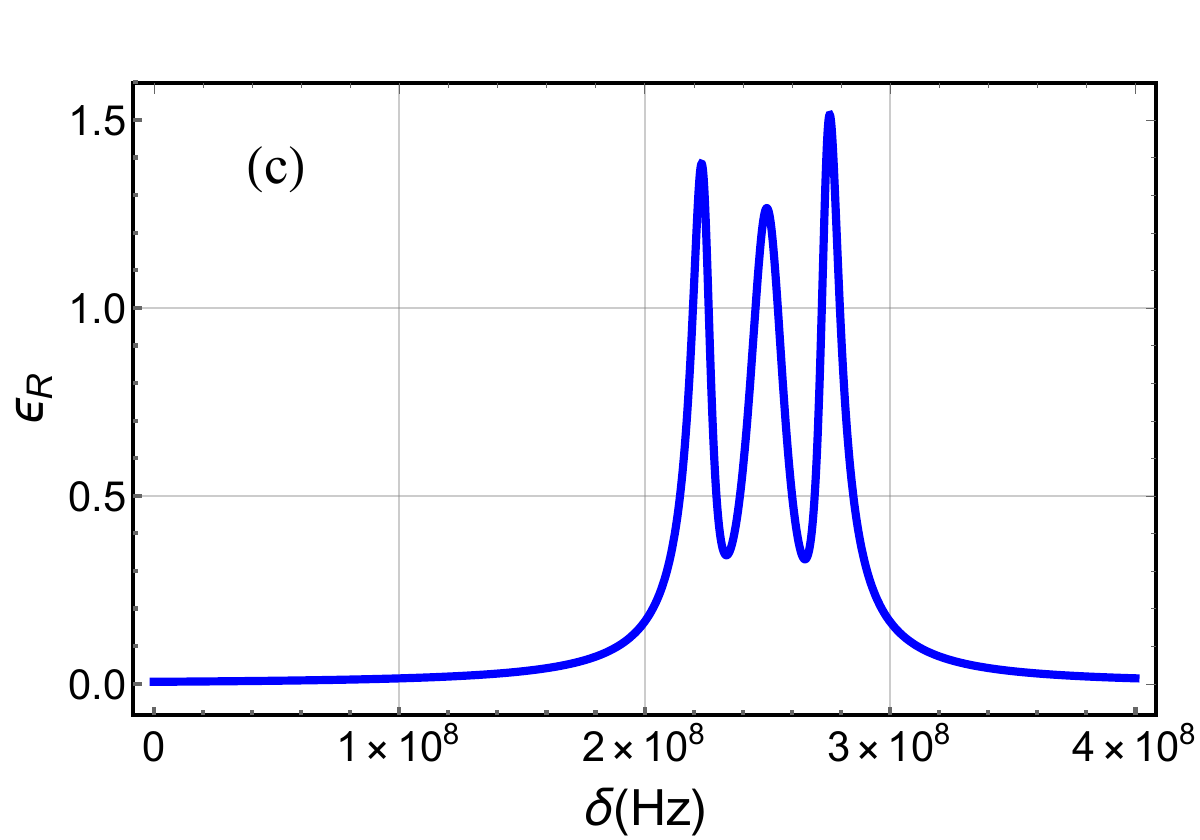}
			\includegraphics[scale=0.35]{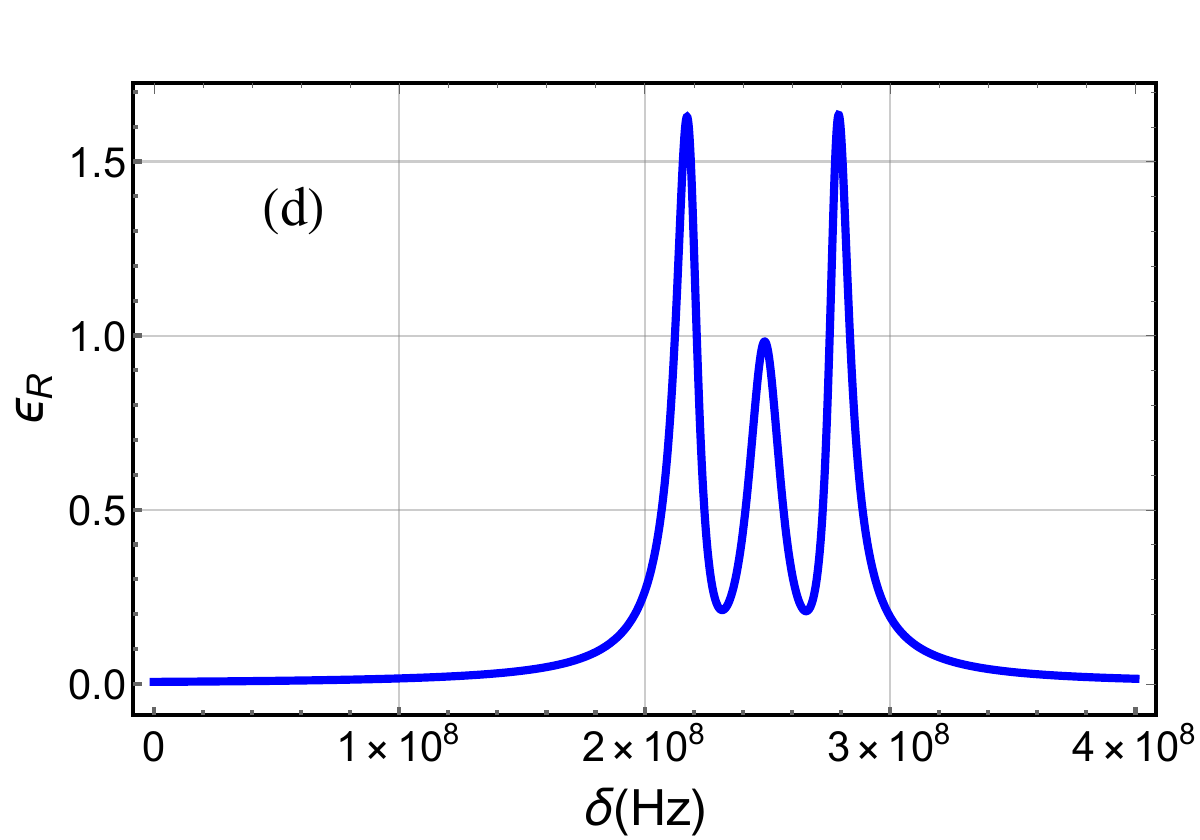}
			\caption{Plot of the absorption $\epsilon_{\text{R}}$ as a function of the detuning $\delta$ for different values of the optomechanical coupling strengths with $G_n/2\pi = 5.6$ MHz. (a) $G_c=0$ MHz, (b) $G_c/2\pi=4$ MHz, (c) $G_c/2\pi=6$ MHz and (d) $G_c/2\pi=8$ MHz. For other parameters, refer to the text}\label{a0}
		\end{center}
	\end{figure}
	
	In figure \ref{a0}, we display the absorption $\epsilon_{\text{R}}$ versus the detuning $\delta$ for different values of the optomechanical coupling strength $G_c$. We observe that there are no signatures of MMIT in the absorption spectra of the output field, as depicted in figure \ref{a0}(a). Moreover, in figure \ref{a0}(b), the double MMIT window appears in the output probe field due to the effective optomechanical coupling $G_c/2\pi = 4$ MHz. In figures \ref{a0}(c) and \ref{a0}(d), the width and the left (right) peaks of the transparency windows increases, at higher values of the effective optomechanical coupling $G_c$. These results clearly prove the existence of a double MMIT window in the output probe field when the cavity optical mode is only coupled to the mechanical mode (phonon).   
	\begin{figure} [h!] 
		\begin{center}
			\includegraphics[scale=0.4]{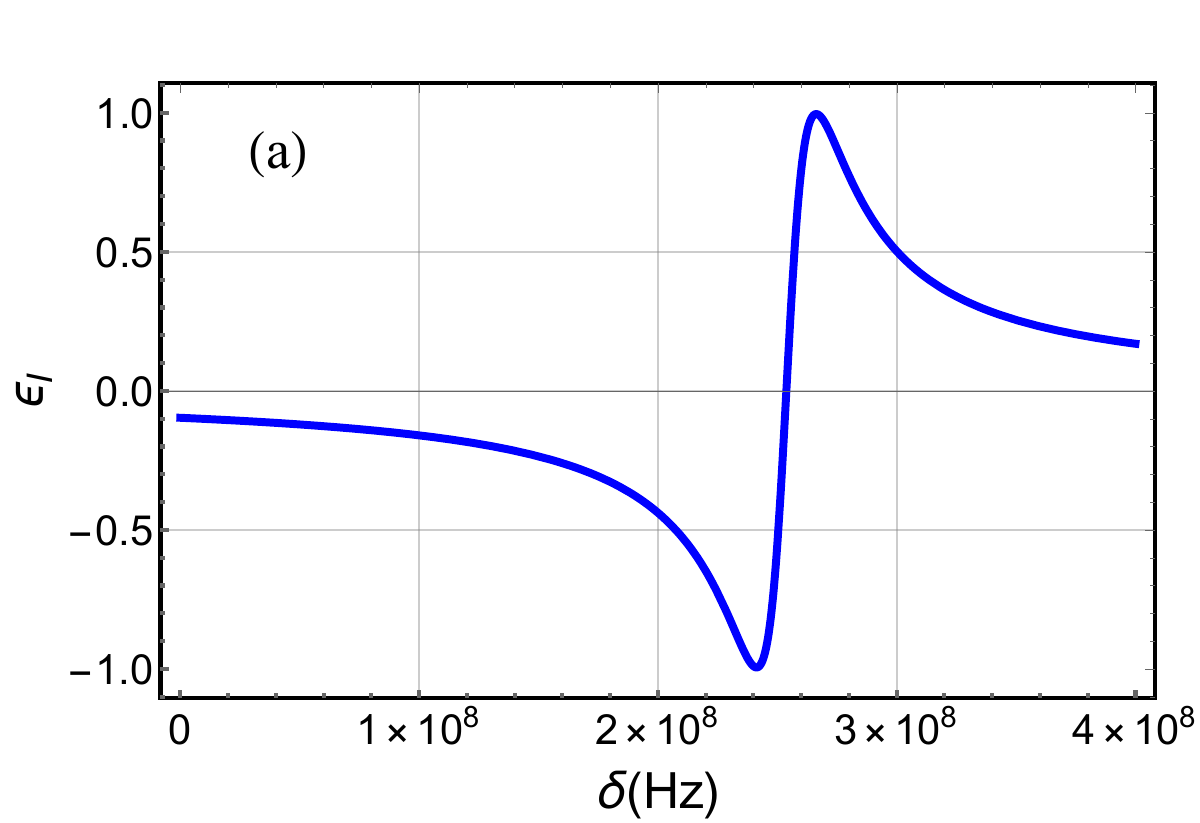}
			\includegraphics[scale=0.4]{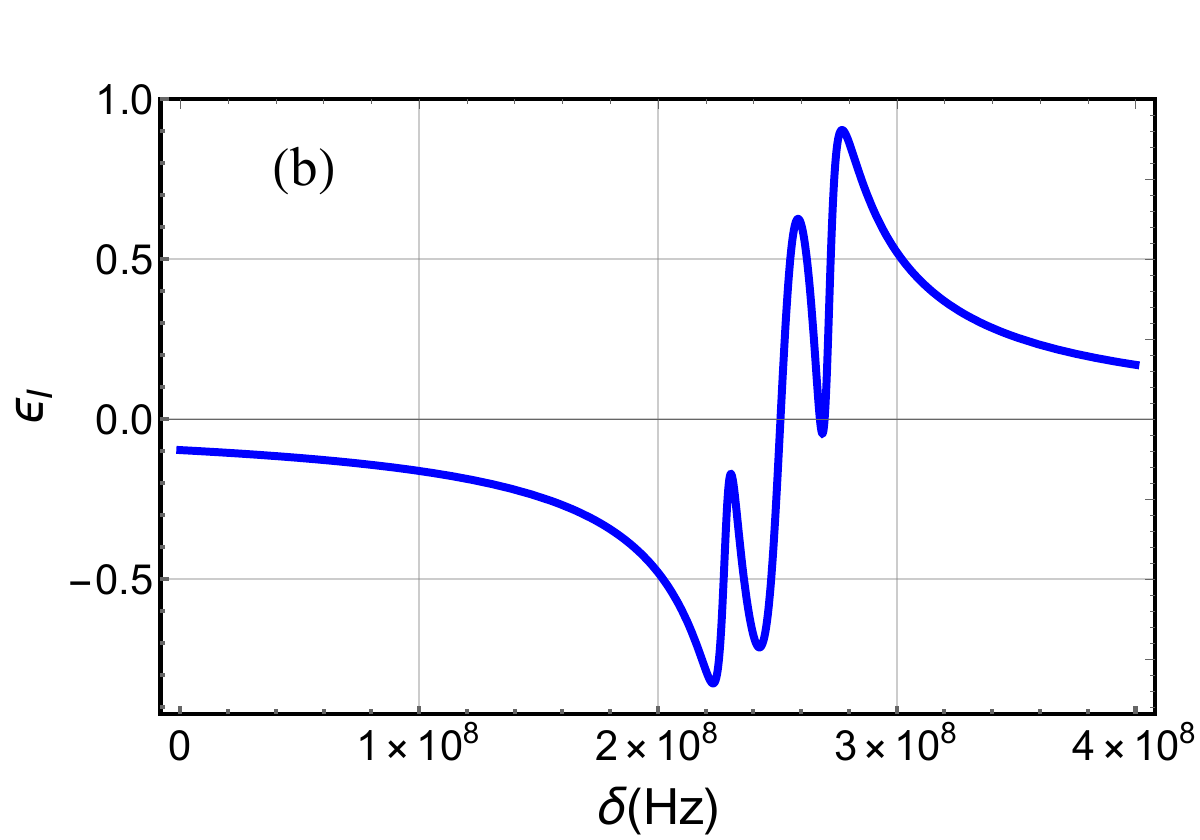}
			\includegraphics[scale=0.4]{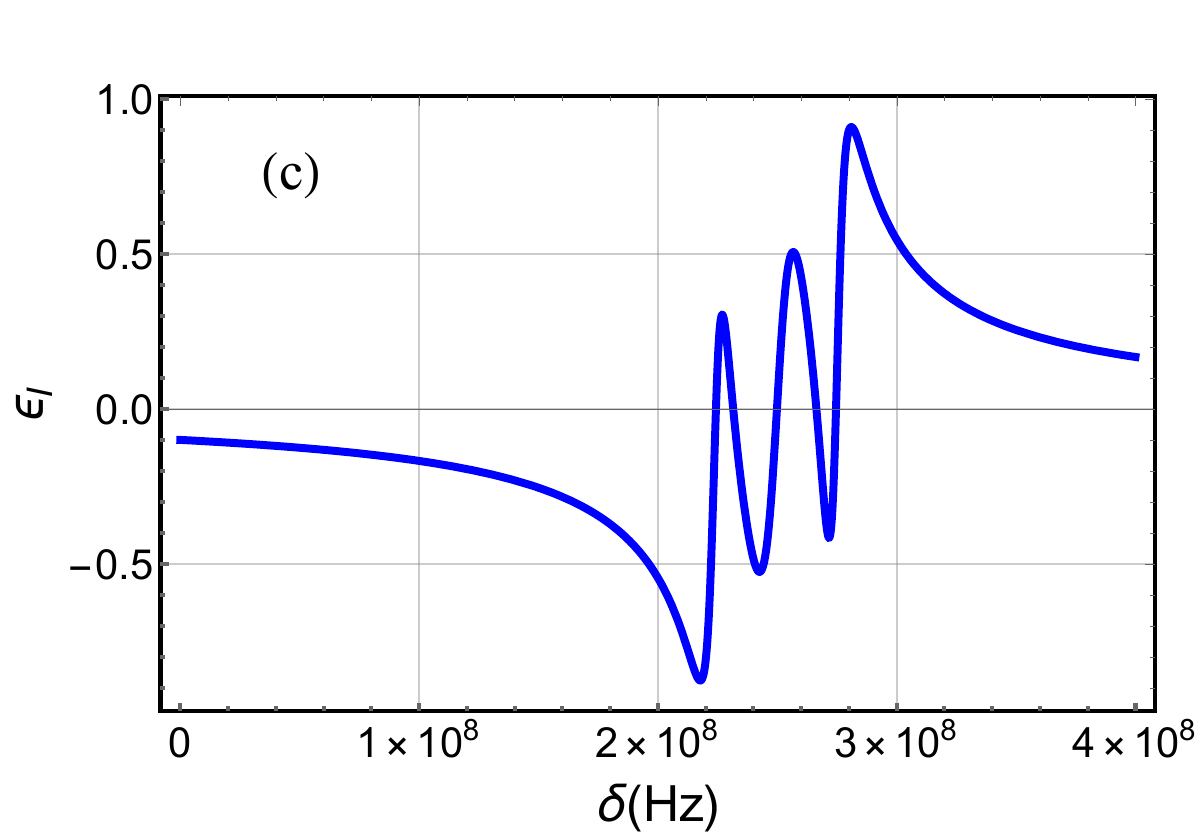}
			\includegraphics[scale=0.4]{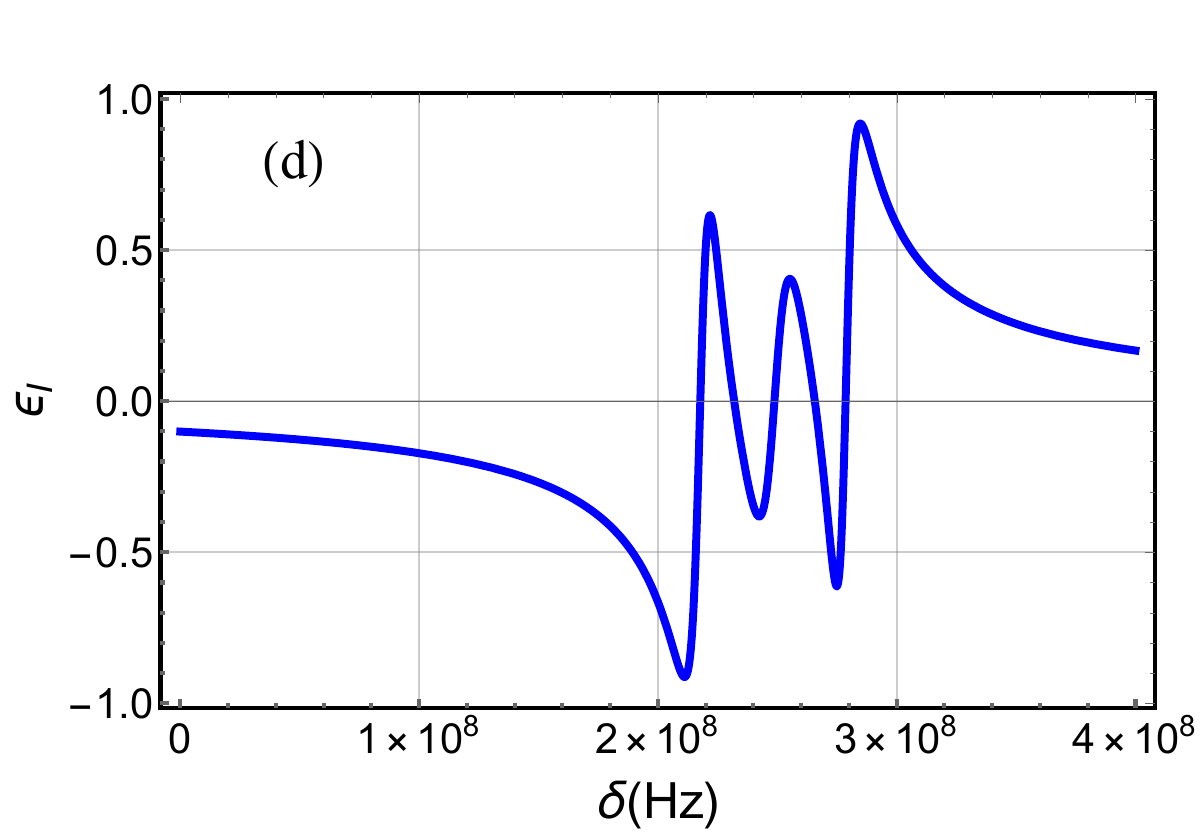}
			\caption{Plot of the imaginary part $\epsilon_{\text{I}}$ of the output field as a function of detuning $\delta$ for different values of the optomechanical coupling strength with $G_n/2\pi = 5.6$ MHz. (a) $G_c=0$ MHz, (b) $G_c/2\pi=4$ MHz, (c) $G_c/2\pi=6$ MHz and (d) $G_c/2\pi=8$ MHz. See text for the other parameters.}\label{a1}
		\end{center}
	\end{figure}
	
	In Figs. \ref{a1}(a)-(d), we plot the dispersion spectrum $\epsilon_{\text{I}}$ of the output field versus the detuning $\delta$ of the probe field for various values of $G_c$ . We note that there are no signatures of MMIT in the dispersion spectra of the output field when the effective optomechanical coupling strength is absent ($G_c=0 $), as shown in Fig. \ref{a1}(a). On the other hand, in Fig. \ref{a1}(b), two MMIT windows appear in the output probe field because the optical mode of the system is now coupled to mechanical mode ($G_c =  4$ MHz). As the effective optomechanical coupling $G_c$ increases, the transparency window  becomes wider, as shown in figures \ref{a1}(c) and \ref{a1}(d).
	
	\section{TRANSMISSION AND TIME DELAY}\label{tree}
	
	In this section, we study the transmission and group delay of the output signal as a function of the detuning $\delta$ with different optomechanical coupling strengths $G_c$. According to Eq. (\ref{0}), the rescaled transmission field associated with the probe field can be written as follows 
	\begin{equation}
		\mathcal{T}=\frac{\epsilon_p-2\kappa_cc_-}{\epsilon_p}
	\end{equation}
	\begin{figure} [h!] 
		\begin{center}
			\includegraphics[scale=0.4]{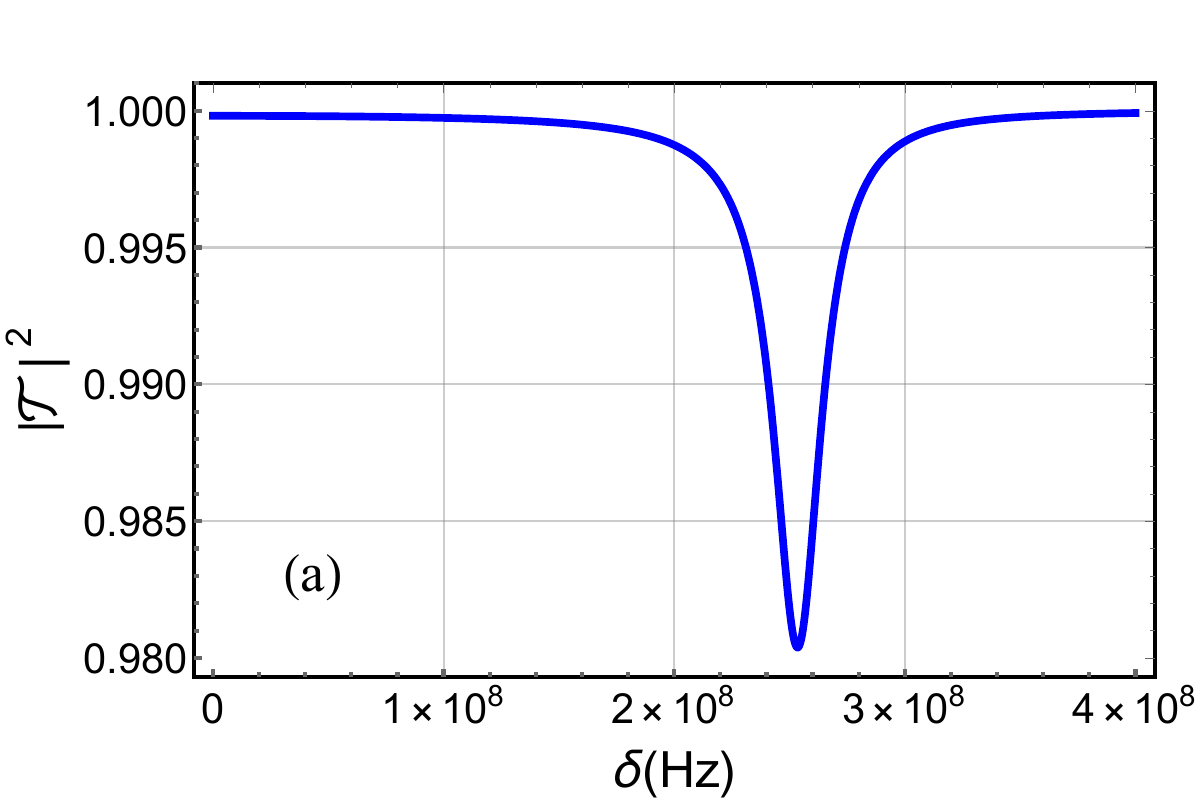}
			\includegraphics[scale=0.4]{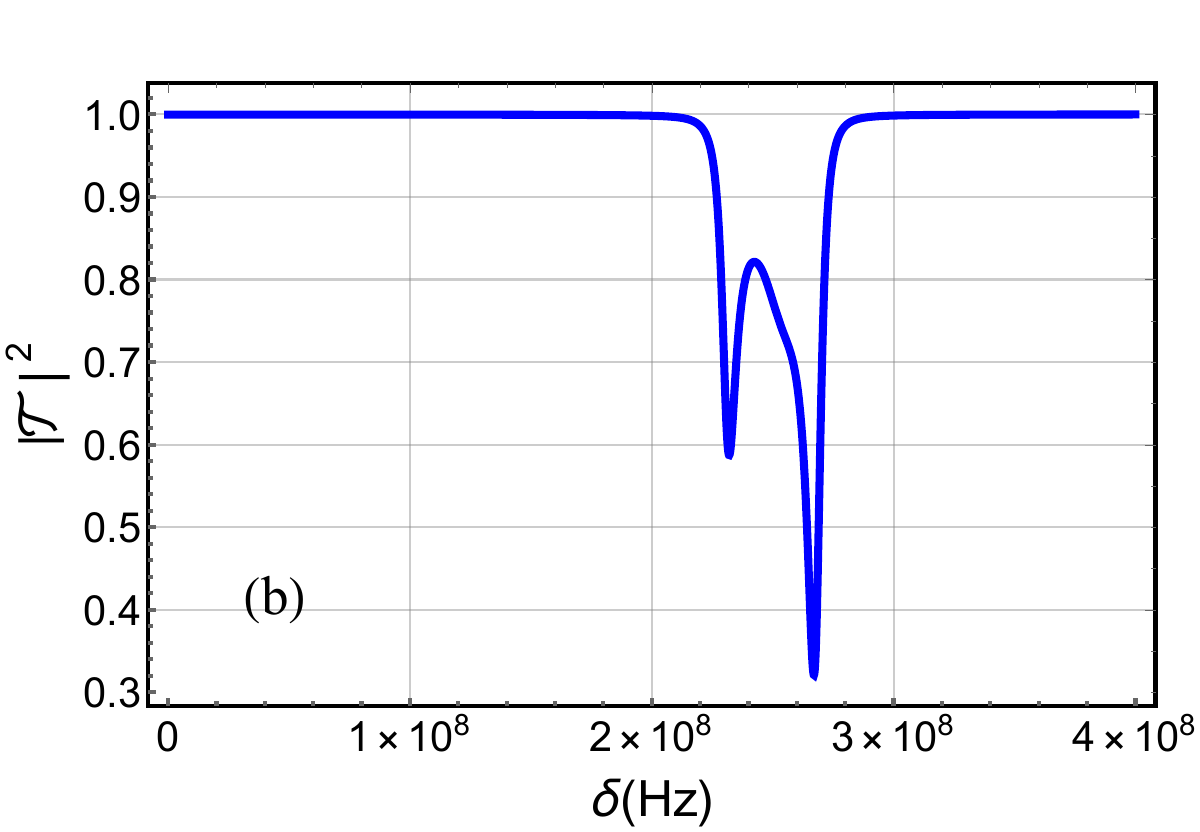}\\
			\includegraphics[scale=0.4]{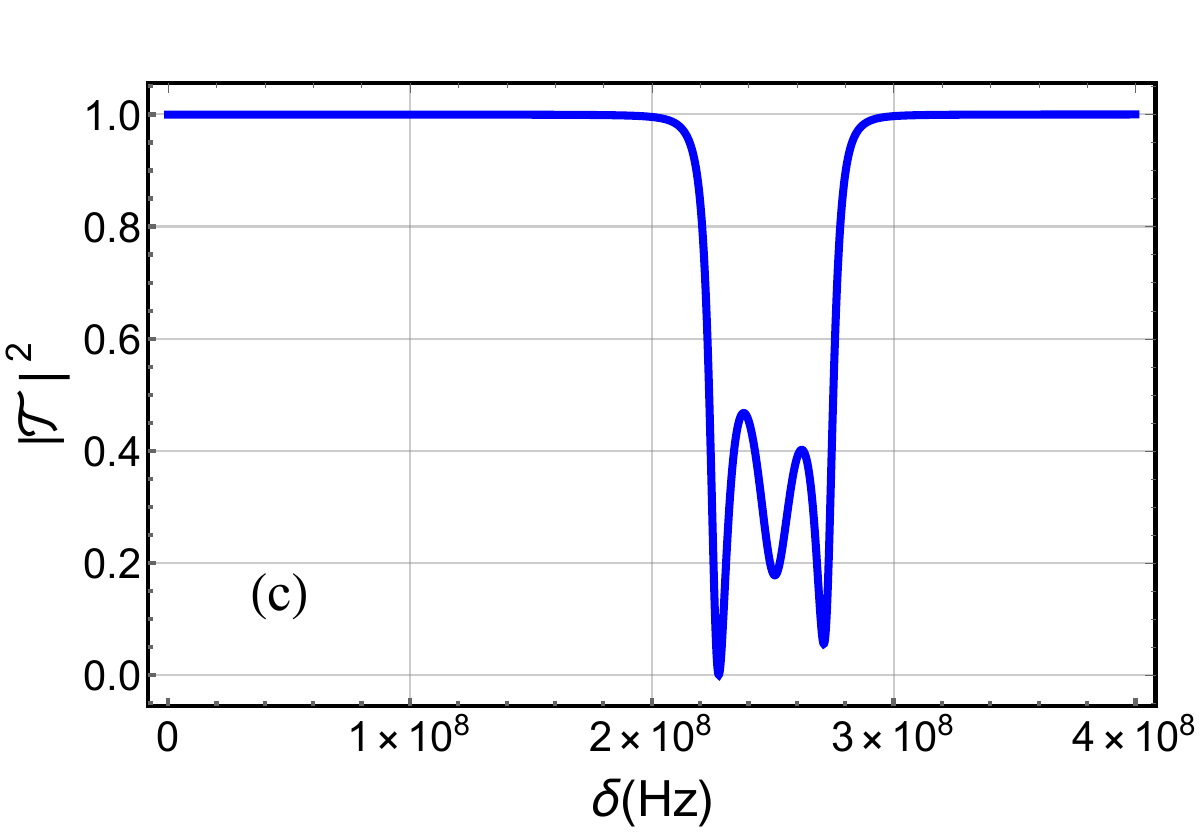}
			\includegraphics[scale=0.4]{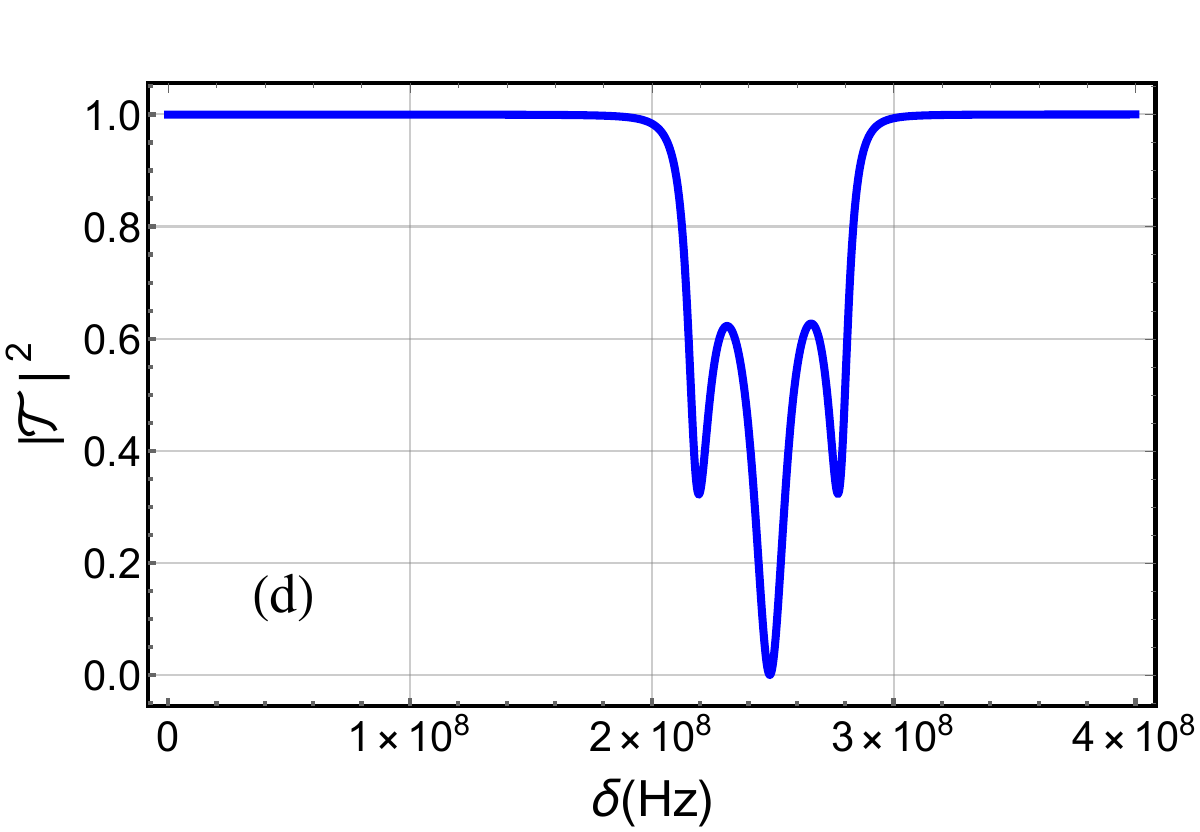}
			\caption{The transmission $|\mathcal{T}|^2$ spectrum as a function of detuning $\delta$ for various values of $G_c$ with $G_n/2\pi = 5.6$ MHz. (a) $G_c=0$, (b) $G_c/2\pi=2$ MHz, (c) $G_c/2\pi=5$ MHz, (d) $G_c/2\pi=8$ MHz. See text for the other parameters.}\label{5}
		\end{center}
	\end{figure}
	
	We plot in figures \ref{5}, the transmission $|\mathcal{T}|^2$ spectrum of the probe field versus the scaled detuning $\delta$, for various values of $G_c$. In Figure \ref{5}(a), $G_c=0$ will not have an effect on the transparent window. It is clear from Fig. \ref{5}(b) that in the presence of optomechanical coupling $G_c/2\pi=4$ MHz, the transmission spectrum of the probe field shows a significant transparency window. Moreover, by increasing the effective coupling strength $G_c$, we observe the appearance of the two small peaks, as implemented in Fig. \ref{5}(c). In addition, Fig. \ref{5}(d) shows that the width of the transparency window increase at higher values of the effective optomechanical coupling $G_c$.

	The phase $\phi$ of the transmitted probe field $\mathcal{T}$ is defined as follows.
	\begin{equation}
		\phi=\text{Arg}[\mathcal{T}]
	\end{equation}
	This phase of the transmitted probe field is associated with the group delay $\tau$ of the output field
	\begin{equation}
		\tau  = \frac{{\partial \phi }}{{\partial {\omega _p}}} = {\mathop{\rm Im}\nolimits} \left[ \frac{1}{\mathcal{T}}\frac{{\partial \mathcal{T}}}{{\partial {\omega _p}}}\right] 
	\end{equation}
	This implies that larger group delays result from quicker phase dispersion and vice versa. Additionally, a negative phase slope indicates a negative group delay or fast light ($\tau < 0$), while a positive phase slope means a positive group delay or slow light ($\tau>0$).
	\begin{figure} [h!] 
		\begin{center}
			\includegraphics[scale=0.4]{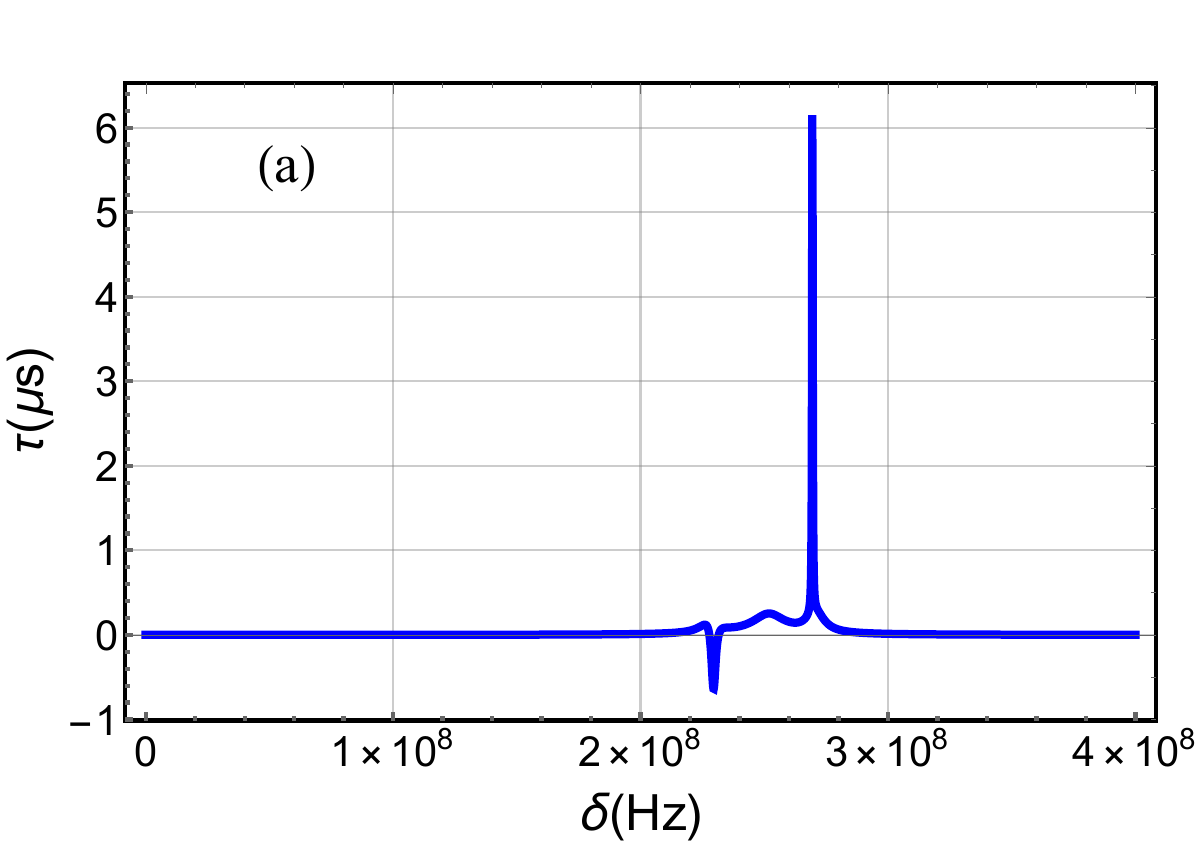}
			\includegraphics[scale=0.4]{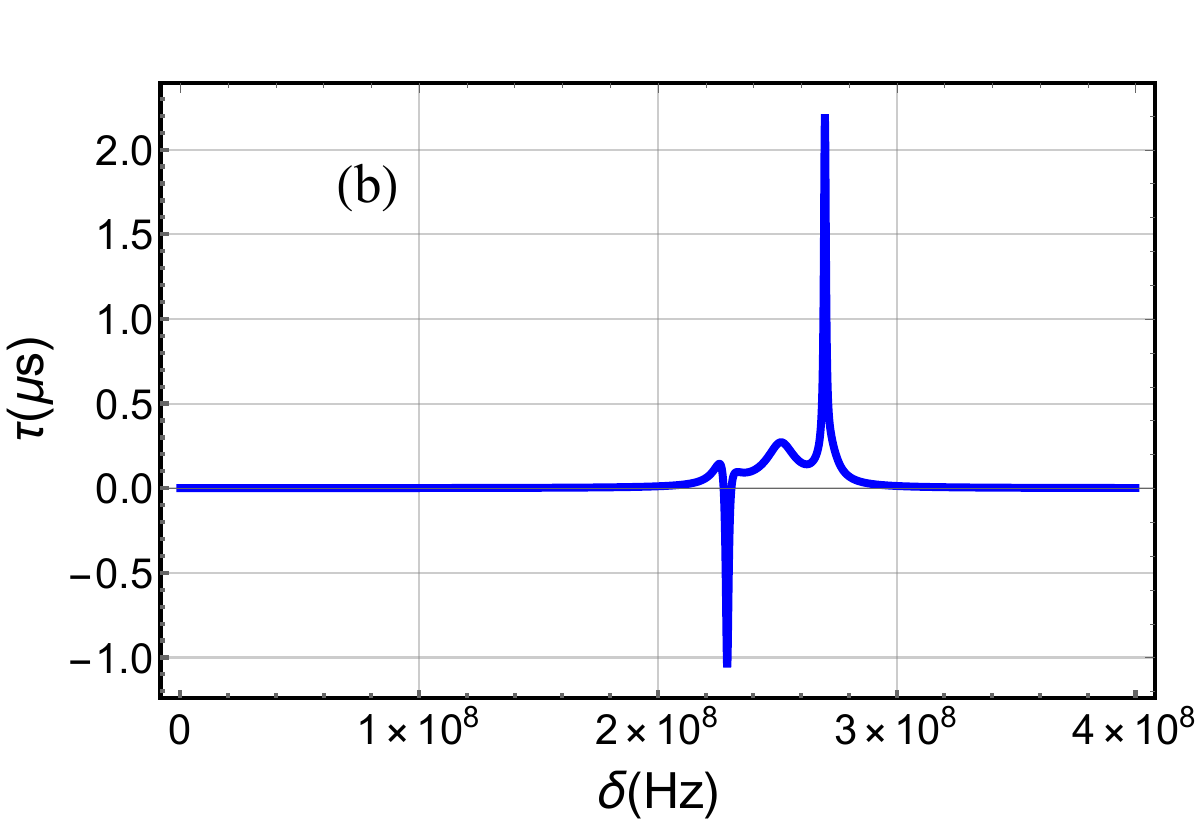}
			\includegraphics[scale=0.4]{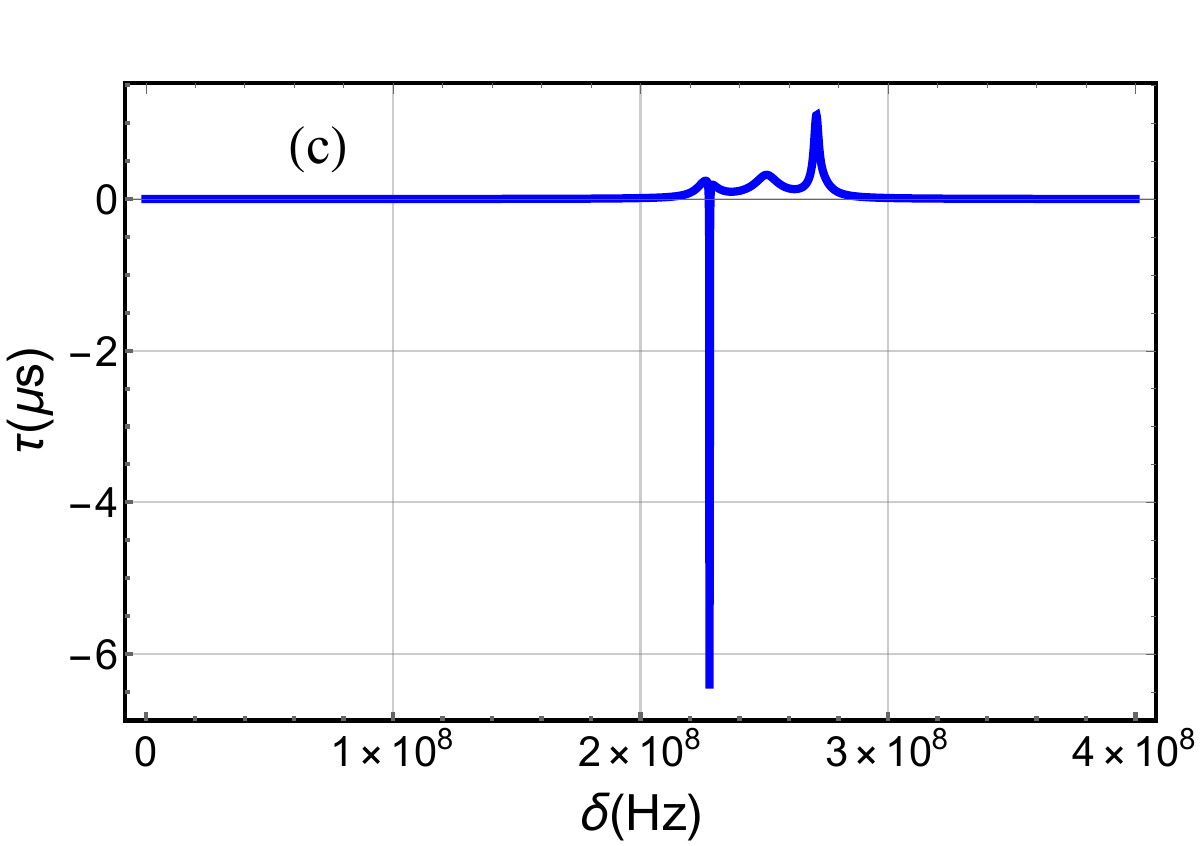}
			\includegraphics[scale=0.4]{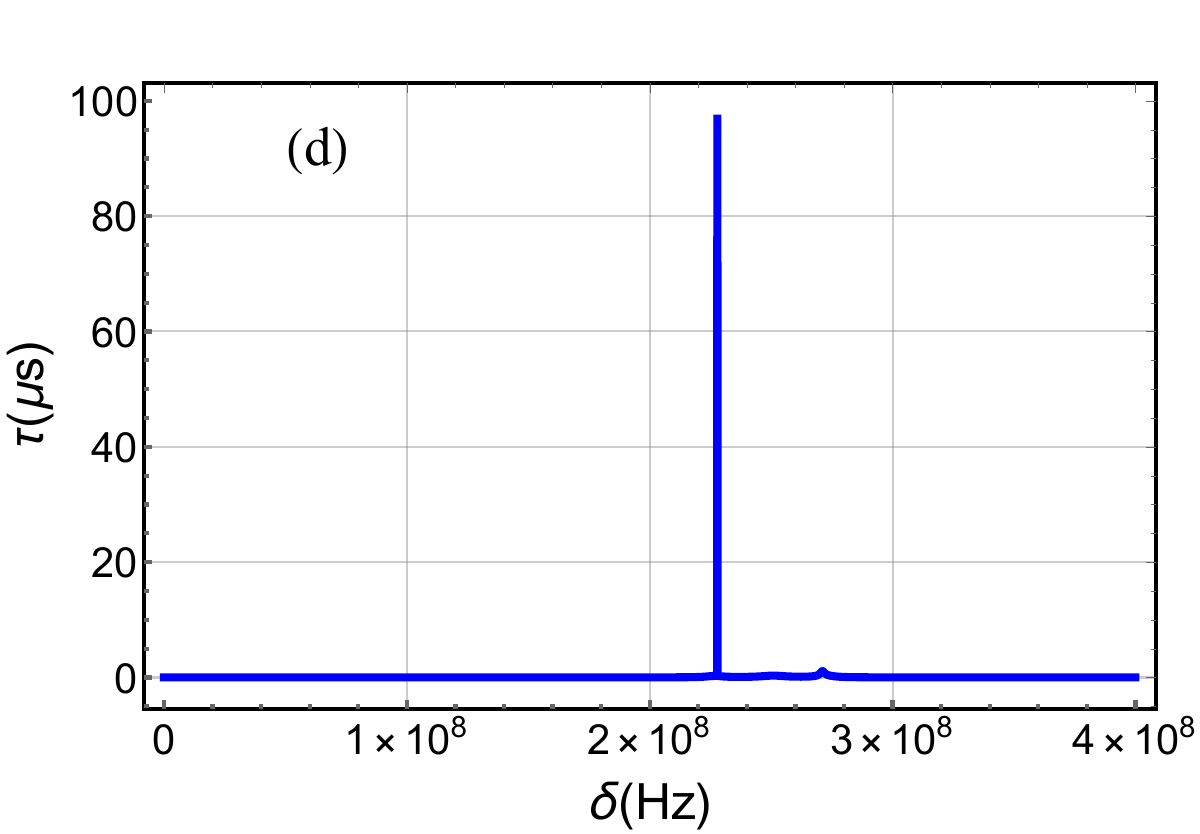}
			\caption{The group delay $\tau$ of the transmitted probe field as a function of detuning $\delta$ for different values of the coupling strength $G_c$ with $G_n/2\pi = 5.6$ MHz. (a) $G_c/2\pi=4$ MHz, (b) $G_c/2\pi=4.30$ MHz, (c) $G_c/2\pi=4.90$ MHz, (d) $G_c/2\pi=4.93$ MHz. See text for the other parameters.}\label{6}
		\end{center}
	\end{figure}
	
	In figure \ref{6}, we plot the group delay of the output field at the probe field frequency versus the detuning $\delta$ for different value of the coupling strength $G_c$. We remark, for $G_c/2\pi=4$ MHz, the group delay at $\delta\approx2.28\times 10^8$ Hz, can be turned to negative value and therefore the output probe field contains fast light. In contrast, the positive group delay at $\delta \approx2.69\times 10^8$ Hz, as shown in figure \ref{6}(a). It is clear from Figs. \ref{6}(b) and \ref{6}(c), by increasing the effective optomechanical coupling $G_c$, the group delay at $\delta\approx2.28\times 10^8$ Hz (slow light) increased and the group delay at $\delta\approx2.69\times 10^8$ Hz (fast light) decreased. This is in comparison with \ref{6}(a). Furthermore, we observe that for $G_c/2\pi=4.93$ MHz, the group delay of the output field is positive, corresponding to the slow light, as depicted in figure \ref{6}(d). 
	
	\section{CONCLUSION}\label{four}
	
In conclusion, we have investigated the magnomechanically induced transparency and slow/fast light in an atom-optomagnomechanical system. We have discussed the absorption, the dispersion, and the transmission and absorption spectra of a weak probe field under a strong control field. Due to the presence of photon-phonon interactions, we have observed optomechanically induced transparency (OMIT), and the phonon-magnon interactions lead to magnomechanically induced transparency (MMIT). We have shown the absence of MMIT as a result of the absence of photon-phonon coupling. We have found two MMIT windows in the output probe spectra due to the presence of photon-phonon and magnon-photon interactions. We investigated slow and fast light propagation conditions that can be modulated by various system parameters. It was shown that, in an opto-magnomechanical system, the group delay enhances by adjusting phonon-photon coupling strength.

\end{document}